\documentclass[nologo,url,11pt,a4paper]{ETHpaper}

\usepackage{amsmath,amssymb}
\usepackage{graphicx}
\usepackage{subfig}
\usepackage[sort&compress]{natbib}
\usepackage{pstricks,pst-node}
\author{Hans-Ulrich Stark, Claudio J.Tessone, Frank Schweitzer}
\address{Chair of Systems Design, ETH Zurich,\\
  Kreuzplatz 5, CH-8032 Zurich, Switzerland}

\title{Slower is faster: Fostering consensus formation by heterogeneous
  intertia}

 \www{\url{http://www.sg.ethz.ch}}
\reference{\emph{ACS - Advances in Complex Systems, vol 11, no 4 (2008) pp. 87-116}}

\begin{document}


\vspace{1cm}

\maketitle

\begin{abstract} 

  We investigate an extension of the voter model in which voters are
  equipped with an individual inertia to change their opinion.  This
  inertia depends on the persistence time of a voter's current opinion
  (ageing).  We focus on the case of only two different inertia values:
  zero if a voter just changed towards a new opinion and $\nu$ otherwise.
  We are interested in the average time to reach consensus, i.e. the
  state in which all voters have adopted the same opinion.  Adding
  inertia to the system means to slow down the dynamics at the voter's
  level, which should presumably lead to a slower consensus formation. As
  an unexpected outcome of our inertial voter dynamics, there is a
  parameter region of $\nu$ where an increasing inertia leads to a faster
  consensus formation.  These results rest on the heterogeneity of voters
  which evolves through the described ageing. In a control setting of
  homogeneous inertia values, we only find monotonously increasing
  consensus times.  In the paper, we present dynamical equations for the
  mean-field case which allow for analytical insight into the observed
  slower-is-faster effect.
\end{abstract}

\section{Introduction}

Decision making means a selection among alternatives.  It is one of the
fundamental processes in economics, but also in social systems. If these
systems consist of many interacting elements -- that we will call
\emph{voters} from now on -- the system dynamics may be described on two
different levels: the \emph{microscopic} level, where the decisions of
the individual voters occur and the \emph{macroscopic} level, where a
certain collective behavior can be observed \citep{fs-wcss-06}.

Based on incomplete information, how does a voter take its decision on a
particular subject? A ``simple'' utility maximization strategy may fail
because in many social situations, for example in public votes, the
private utility cannot easily be quantified, i.e. voters do not exactly
know about it.  So, voters have to involve supplemented strategies to
take their decisions.  In order to reduce the risk of making the wrong
decision, it seems to be appropriate just to copy the decisions of
others. Such an \emph{imitation} strategy is widely found in biology, but
also in cultural evolution. Different species, including humans, imitate
the behavior of others of their species to become successful or just to
adapt to an existing community~\cite{dugatkin01}.

In order to understand the intrinsic properties of systems comprising
many such individuals, a number of models have been developed that take
the spread of opinions as sample application. Early approaches in the
social sciences showed that the existence of positive social influence
(i.e. imitation behavior) tends to establish homogeneity (i.e. consensus)
among individuals~\cite{french56fts,abelson64mmd}. The ``voter model''
(VM), rigorously defined by ~\citet{ligget1995}, confirms these results.
Other works showed that selection of interaction partners (``bounded
confidence''~\cite{deffuant00mba,hegselmann02oda}) can lead to stable
diversity of opinions, even when considering positive social influence.

Here, we focus on the VM -- a paradigmatic model to simulate such
imitation behavior. Because of its simplicity, it allows for many
analytical calculations ~\cite{ligget1995,redner2001} and, therefore,
serves a comprehensive understanding of the dynamics involved.
Application areas of the VM range from coarsening phenomena
\cite{dornic2001}, spin-glasses \cite{ligget1995,fontes2001}, species
competition~\cite{ravasz04,chave2001}, and opinion
dynamics~\cite{holyst2001}. Among the most prominent properties of the
VM, the conservation of magnetization has extensively been studied
\cite{frachenbourg1996,castellano2003,suchecki2005} and compared to other
prototypical models, such as the Ising Model with Kawasaki dynamics
\cite{gunton1983}.

Based on the VM, investigations were conducted to study interesting
emergent phenomena and relevant applications. Such works comprise the
possibility of minority opinion spreading~\cite{galam,sanmiguel},
dominance in predator-prey systems~\cite{ravasz04}, forest growth with
tree species competition~\cite{chave2001}, and the role of bilingualism
in the context of language competition~\cite{xavi}.  The question of
consensus times and their scaling for different system characteristics
was particularly addressed in several
studies~\cite{ligget1995,redner2001,sood2005,castellano2003,maxi2}.

In this paper, we study a modified version of the VM introduced
recently~\cite{PRL}.  There, we assume that an individual voter has a
certain inertia $\nu_{i}$ to change its opinion. $\nu_{i}$ increases with
the persistence time $\tau_{i}$ which is the time elapsed since the last
change of opinion.  The longer the voter already stays with its current
opinion, the less it may be inclined to change it in the next time
step. We show that this slowing-down of the dynamics at the microscopic
level of the voters can lead to an accelerated formation of consensus at
the macroscopic level. In this paper, we extend the previous results by
presenting a reduced description of the model which bases on only two
levels of inertia.  We show that this reduction still explains the origin
of the faster consensus formation and thus complement the results
presented in~\cite{PRL}. Moreover, we emphasize the relevance of our
approach for the research of social dynamics.

At difference with the standard VM, our extension considers the current
opinion of voters as an important decisive factor.  The voters do not
only act based on the frequencies in their neighborhood, but take their
own current opinion into particular account. This general idea can also
be compared to the models of continuous opinion dynamics
(see~\cite{deffuant00mba,hegselmann02oda,lorenz07cod}): already in the
basic continuous models, the current opinion of a deciding individual is
of high importance. More precisely, it is as decisive as the average
opinion in the considered neighborhood because the updated opinion is the
average of both.  The concept of bounded confidence emphasizes this
importance because individuals do only approach opinions that are not too
far away from their own current one.  Therefore, bounded confidence can
also be interpreted as a kind of inertia that tend to let individuals
keep their own current opinion.  However, the parameter regulating the
confidence interval is generally kept constant in time whereas, in our
model, individuals change their decision behavior dependent on their
history.

Our new parameter $\nu$, that reduces the probability of state changes in
the VM, can have different interpretations in the various fields of
application of the VM: it may characterize molecules that are less
reactive, the permanent alignment of spins in a magnet, etc. In
economics, changes may be discarded due to transition- or ``sunk'' costs.
In social applications, there are at least two interpretations for the
parameter $\nu$: (i) within the concept of \emph{social inertia}, which
deals with a habituation of individuals and groups to continue their
behavior regardless of possible advantages of a change, (ii) to reflect a
(subjective or objective) conviction regarding a view or an opinion.
Originally, the latter point served as a motivation for us to study the
implications of built-in conviction in a simple imitation model like the
VM. Will the systems, dependent on the level of conviction, still reach a
consensus state, or can we observe the segregation of opinions? How does
the ordering dynamics and the emergent opinion patterns look like?

Our investigations focus on the average time to reach consensus, i.e. the
number of timesteps the system evolves until it reaches an equilibrium
state in which all voters have the same opinion. Taking into account the
inertia introduced to the VM, we would assume that the time to reach
consensus shall be increased because of the slowed-down voter dynamics.
Counter-intuitively, we find that increasing inertia in the system can
{\em decrease} the time to reach consensus.  This result resembles the
``faster-is-slower'' effect reported in a different context
by~\citet{helbing}. In their work on panic situations, they explain why
rooms can be evacuated faster if people move slower than a critical value
through the narrow exit door. When individuals try to get out as fast as
they can, this results in clogging effects in the vicinity of the door,
which decreases the overall evacuation speed. Note that although the
phrase ``slower-is-faster'' is appropriate for both findings, our effect
has to be clearly distinguished from the one described by Helbing
\emph{et al.}.  In their generalized force model, an individual increase
in the desired velocity would have a contrary effect on the microscopic
level, i.e. all individuals would get slower and thereby the macroscopic
dynamics would be decelerated. In our case, microscopic changes produce
the counter-intuitive effect only on the macroscopic level.

This paper is organized as follows: In the following Section~\ref{model},
we introduce the model. In Section~\ref{results} we present simulation
results of our main finding, the ``slower-is-faster'' effect on reaching
consensus through inertial voters. Section~\ref{analysis} investigates
deeper, under which circumstances the effect can be observed and
introduces a theoretical framework, that allows to understand the
phenomenon. Finally, the conclusions are drawn in
Section~\ref{conclusion}.

\section{The model\label{model}}
\subsection{The standard voter model}

In the original voter model~\cite{holley75,ligget1995,dornic2001}, $N$
voters are positioned at the sites of a regular, $d-$dimensional lattice,
the topology which defines the number of neighbors for each voter.  Every
voter has one of two possible opinions $\sigma_i=\pm1$.  A timestep
consists of $N$ update events in each of which one voter is picked at
random and adopts the opinion of one of the voters he is connected to.
Thus, the probability that voter $i$ adopts opinion $\sigma$, that we
will denote $W_i^{V}(\sigma)$, is equal to the density of opinion
$\sigma$ in its neighborhood.  Hence,
\begin{equation} 
  W_i^V(\sigma,t) \equiv W_i^{V}(\sigma|\sigma_i,t)=\frac{1}{2}\left(1+\frac{\sigma}{k}\sum_{j\in
      \lbrace i \rbrace }\sigma_j (t)\right),\label{lvm}
\end{equation}
where $k$ is the number of neighbors each voter has, and $\lbrace i
\rbrace$ is the set of its neighbors. Note that this equation can also be
applied to networks of different topology, as we will do later on.

The dynamics is a fluctuation driven process that, for finite system
sizes, ends up in one of two absorbing states, i.e., consensus in one of
either opinions.  The time to reach consensus, $T_\kappa$, depends on the
size of the system and the topology of the neighborhood network. For
regular lattices with dimension $d=1$, $T_\kappa\propto N^2$, for
$d=2,T_\kappa\propto \ln N$, and for $d>2$, $T_\kappa\propto N$.  A
critical dimension $d=2$ was found, below which the system coarsens. For
any dimension larger than 2, the system can get trapped in disordered
configurations in infinite systems~\cite{slanina2003}.

Let $P_\sigma(t)$ be the global density of voters with opinion $\sigma$
at time $t$. The average opinion of the system (also called
``magnetization'' analogous to studies of spin systems in physics) can be
computed as
\begin{equation}
M(t) = P_+(t) - P_-(t). \label{magnet}
\end{equation}
The order parameter, most often used in the voter model, is that of the
average interface density $\rho$. It gives the relative number of links
in the system that connect two voters with different opinions and can be
written as
\begin{equation}
  \rho(t) =  \frac 1 4 \sum_{i} \sum_{j\in \lbrace
    i \rbrace } \big( 1-\sigma_i(t)\, \sigma_j(t) \big). 
\label{density}
 \end{equation}
 In the mean field limit, that we will study in more detail later on, we
 assume that the change in the opinion of an individual voter only
 depends on the average densities of the different opinions in the whole
 system. Therefore, we replace the local densities Eq.~(\ref{lvm}) by
 global ones, which leads to the adoption probabilities $W^V(\sigma |
 -\sigma,t) = P_\sigma(t)$.  For the macroscopic dynamics, we can compute
 the change in the global density of one opinion as
\begin{eqnarray}
  \label{vm_mf}
  P_\sigma(t+1)-P_\sigma(t) &=& W^V(\sigma | -\sigma,t)P_{-\sigma}(t) -
W^V(-\sigma|\sigma,t)P_{\sigma}(t)\nonumber\\
  &=& P_\sigma(t)P_{-\sigma}(t) -  P_{-\sigma}(t)P_{\sigma}(t)\nonumber\\
  &\equiv& 0,
\end{eqnarray}
i.e. the density of each opinion is conserved for every state of the
system. In the simulations, consensus is reached by finite-size
fluctuations only.

\subsection{The voter model with social inertia}

Different from the standard VM described above, we consider that voters
additionally are characterized by a parameter $\nu_i$, an inertia to
change their opinion. This extension leads us to the \emph{inertial
  voter model}, in which we have to distinguish between the probability
that voter $i$ changes its opinion
\begin{equation}
  \label{ivm}
  W_i(-\sigma_i|\sigma_i,\nu_i)=(1-\nu_i)\,W_i^{V}(-\sigma_i) 
\end{equation}
and the complementary probability of sticking to its previous opinion
$W_i(\sigma_i|\sigma_i,\nu_i)=1-W_i(-\sigma_i|\sigma_i,\nu_i)$. In this
setting, $\nu_i$ represents the strength of ``conviction'' that voter $i$
has regarding its opinion.

We consider that the longer a voter has been keeping its current opinion, the
less likely he will change to the other one. For the sake of simplicity, we
consider that the inertia grows with the persistence time as
\begin{equation}
  \label{reluct}
  \nu_i(\tau) =
\left\{ \begin{array}{lrr}
\nu_{0}, & \,\,\hbox{if} & \tau=0\\
 \nu, & \,\,\hbox{if} & \tau>0
\end{array}
\right. .
\end{equation}
At time $t=0$, and in every timestep after voter $i$ has changed its
opinion, the persistence time is reset to zero, $\tau_i=0$, and the
inertia has the minimal constant value $\nu_{0}$.\footnote{ Note that the
  results of this paper are qualitatively robust against changes in the
  concrete function $\nu_i(\tau)$. For more details on this
  see~\cite{PRL}.} Whenever a voter keeps its opinion, its inertia
increases to $\nu$.  We will study two distinct scenarios later on: (i)
fixed social inertia where $ \nu_{0} = \nu$ is a constant value for all
voters.  (ii) $\nu_{0} < \nu$, a scenario in which inertia grows for
larger persistence times.

It would be expected that including inertial behavior in the model would
invariably lead to a slowing-down of the ordering dynamics. We will show
that, contrary to this intuition, these settings can lead to a much faster
consensus.

\section{Numerical Results\label{results}}

We performed extensive computer simulations in which we investigated the
time to reach consensus, $T_\kappa$, for systems of $N$ voters.  We used
random initial conditions with equally distributed opinions and an
asynchronous update mode, i.e. on average, every voter updates its opinion
once per timestep. The numerical results correspond to regular
$d-$dimensional lattices (von-Neumann neighborhood) with periodic
boundary conditions, and small-world networks with an homogeneous degree
distribution.

\subsection{Fixed social inertia}

\begin{figure}[ht!]
  \centering
\includegraphics[width=13cm]{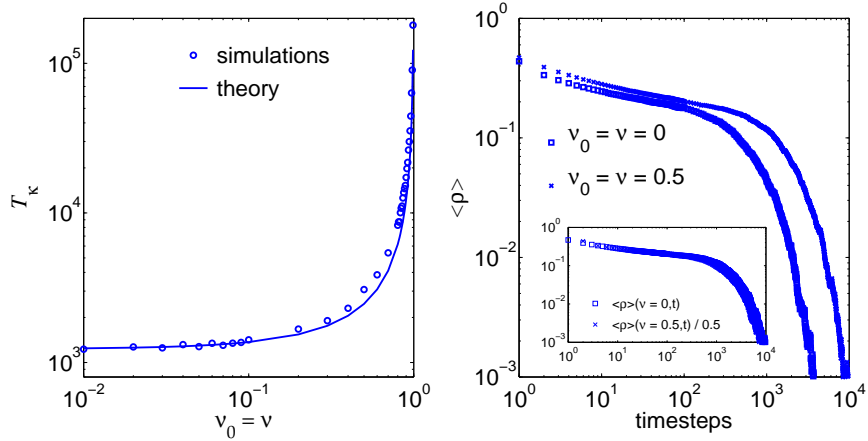}
\caption{ (Left panel) Average times to consensus $T_\kappa$ in the voter
  model with a fixed and homogeneous inertia value $\nu_0=\nu$.  The line
  corresponds to the theoretical prediction $T_\kappa(\nu) = T_\kappa(\nu
  = 0) / (1-\nu)$, whose details are given in the text.  (Right panel)
  Comparison of the development of the average interface density $\rho$
  in the voter model and the model with fixed inertia. The curves
  correspond to the mean values obtained out of 500 realizations.  (Right
  panel, inset) Collapse of the curves where the time scale has been
  rescaled according to $t \to t / (1-\nu)$. The system size is
  $N=30\times 30$ in both panels and the voters are placed in a
  two-dimensional regular lattice. }
  \label{fixed}
\end{figure}

We first consider the case of a \emph{fixed} and \emph{homogeneous}
inertia value $\nu_0=\nu$. In the limit $\nu \to 0$, we recover the
standard VM, while for $\nu = 1$ the system gets frozen in its initial
state. For $0 \leq \nu < 1$, the time to reach global consensus shall be
affected considerably, i.e. the system will still always reach global
consensus, but this process is decelerated for higher values of $\nu$.
This can be confirmed by computer simulations which assume a constant
inertia equal for all voters (see left panel in Fig.~\ref{fixed}).

In the right panel of Fig.~\ref{fixed}, we depict the evolution of the
interface density $\rho$ 
for both the standard VM and the inertial VM with $\nu_0 = \nu =0.5$.
Differences between these cases can be seen in the very beginning and at
about $10^{3}$ time steps, right before the steep decay of disorder in
the system. There, the ordering process is slower than in the VM without
inertia. The distributions of $T_\kappa$ at different $\nu$-values are
very similar and show the log-normal like form known from the standard VM
($\nu=0$).

This behavior can be well understood by analyzing Eq.~(\ref{ivm}). It can
be seen that the ratio between the opinion changes in the standard VM and
the inertial VM is given by
$W(-\sigma|\sigma,\nu)/W^V(-\sigma|\sigma,0)=(1-\nu)$. Consequently, it
is possible to infer that the characteristic time scale for a VM with
fixed inertia will be rescaled as $t \to t_{V}/(1-\nu) $. As can be seen
in Fig.~\ref{fixed}, there is good agreement between this theoretical
prediction and computer simulations in both: the average time to
consensus (see panel (a)), and the time evolution of the interface
density (inset of panel (b)).

\subsection{Evolving social inertia}

\begin{figure}[ht]
\vspace{20pt}
  \centering
   \includegraphics[width=13cm]{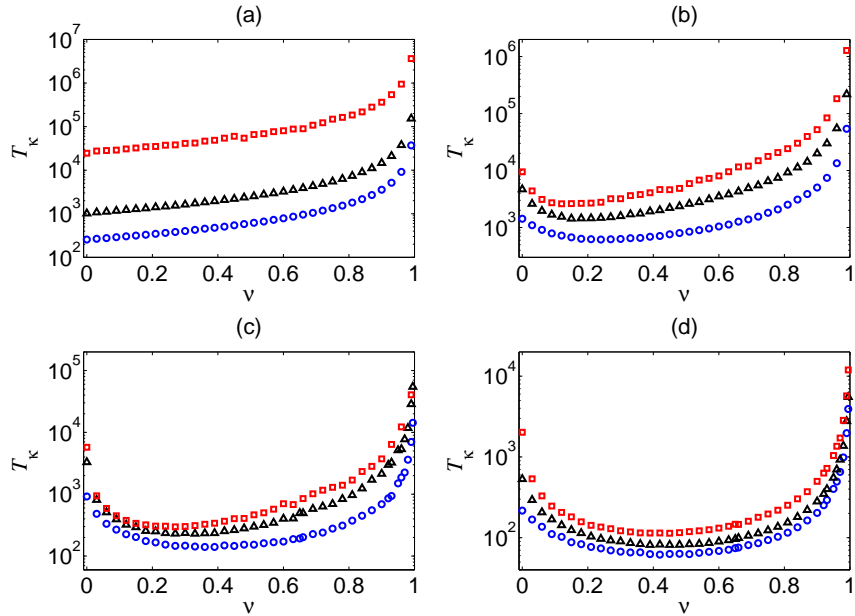}
   \caption{Average time to reach consensus $T_\kappa$ as a function of
     the maximum inertia value $\nu$. Panels (a), (b), (c) and (d) show
     the results for different system sizes in one-, two-, three-, and
     four-dimensional regular lattices, respectively. The results are
     averaged over $10^4$ realizations. The system sizes for the
     different panels are the following: (a) $N=50$ ($\circ$), $N=100$
     ($\triangle$), $N=500$ ($\square$); (b)
     $N=30^2$ ($\circ$), $N=50^2$ ($\triangle$), $N=70^2$ ($\square$); (c)
$N=10^3$
     ($\circ$), $N=15^3$ ($\triangle$), $N=18^3$ ($\square$); (d)
     $N=4^4$ ($\circ$), $N=5^4$ ($\triangle$), $N=7^4$ ($\square$).}
  \label{ttc}
\end{figure}

We now turn our attention to the case where the individual inertia values
evolve with respect to the persistence time according to
Eq.~(\ref{reluct}). Without loss of generality, we fix $\nu_0=0$. Other
choices simply decelerate the overall dynamics as described in the
previous subsection. Note that increasing $\nu$ increases the level of
social inertia in the voter population.  Fig.~\ref{ttc} shows the average
time to reach consensus as a function of the parameter $\nu$, namely the
maximum inertia value reached by the voters when the system is embedded
in regular lattices of different dimensions.  In Fig.~\ref{ttc}, it is
apparent that, for lattices of dimension $d \geq 2$, the system exhibits
a noticeable reduction in the time to reach consensus for intermediate
values of the control parameter $\nu$. We observe that there is a
critical value of $\nu$ such that the average consensus time reaches a
minimum. Especially compared to the results of the previous section, this
result is against the intuition that a slowing-down of local dynamics
would lead to slower global dynamics.  Furthermore, it is also apparent
that the larger the dimension of the lattice, the more pronounced is the
phenomenon is.  Fig.~\ref{ttc}(a) shows the results for a one-dimensional
lattice, where the phenomenon is not present at all.  For this network
topology, it is found that all the curves collapse according to a scaling
relation $T_{\kappa}(\nu,N) = T_\kappa(\nu) / N^2 $.

\begin{figure}[t!]
  \centering
  \includegraphics[width=8cm]{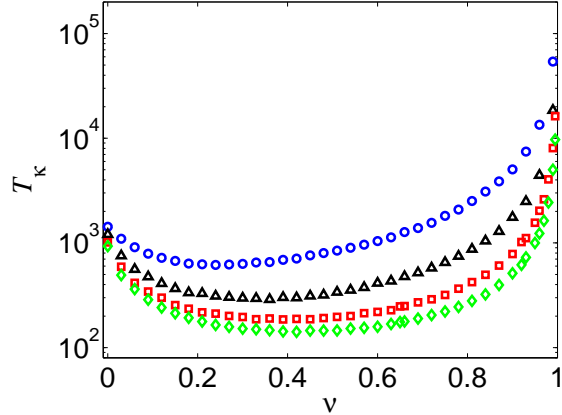}
  \caption{Dependence of the average time to consensus on the control
    parameter $\nu$. The symbols represent different rewiring
    probabilities $\omega$, when the network topology is a small-world one.
    The curves corresponds to $\omega=0$ ($\circ$), $\omega=0.03$
($\triangle$),
    $\omega=0.1$ ($\square$), and $\omega=0.9$ ($\diamondsuit$).}
  \label{sw}
\end{figure}

In Fig.~\ref{sw}, we plot $T_\kappa$ as a function of the maximum inertia
value $\nu$ for different small-world networks~\cite{watts1998}. Starting
with a two-dimensional regular lattice, two edges are randomly selected
from the system and with probability $\omega$ their end nodes are
exchanged~\cite{maslov2003}. With this procedure, the number of neighbors
remains constant for every voter.  It can be seen that the phenomenon of
lower consensus times for intermediate inertia values is also present in
small world networks.  Furthermore, increasing the rewiring probability
$\omega$ leads to larger reductions of the consensus times at the optimal
value $\nu_c$. This implies that the formation of spatial configurations,
such as clusters, is not the origin of this slower-is-faster effect.

Finally, we show the results on a fully-connected network, i.e.~where
every voter has $N-1$ neighbors. The results are shown in Fig.~\ref{num}.
As can be seen, the time to reach consensus is significantly decreased
for intermediate values of $\nu$.

\section{Analytical approach}\label{analysis}

\begin{figure}[ht]
  \centering
  \includegraphics[width=9cm]{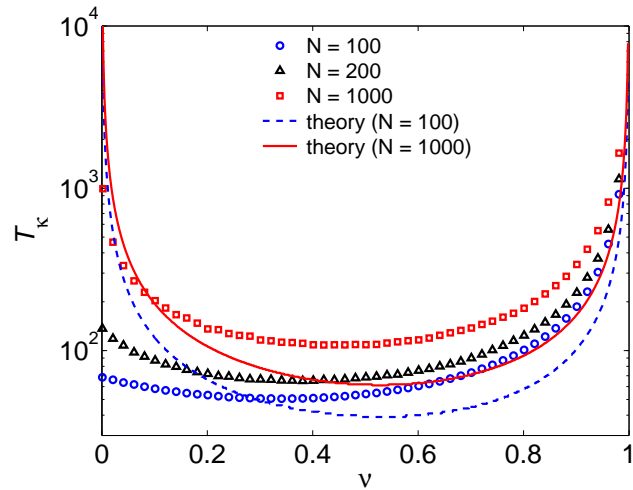}
  \caption{Averaged times to consensus $T_\kappa$ as a function of the
    value of social inertia $\nu$. Symbols show the simulation results
    for different system sizes, intersected lines the results of the
    theoretical estimation. A fully connected network of voters was
    used.}
  \label{num}
\end{figure}

As already mentioned, the 
results of Figs.~\ref{sw} and \ref{num} indicate that the spatial
clustering plays no important role for the voters' ageing\footnote{By
  ageing we mean the possibility to build up higher persistence times
  that in turn lead to increasing inertia values.} and, therefore, for
the qualitative behavior observed. This finding allows for a quantitative
approach to the dynamics in the mean-field limit, i.e., we now use the
{\em global} densities of opinions to calculate the probability
$W_i(-\sigma_i|\sigma_i,\nu_i)$ in Eq.~(\ref{ivm}).

Let us first introduce $p_l^\sigma(t)$ as the fraction of voters with
opinion $\sigma$ and inertia state $l$. I.e.~$l=1$ if they are inertial
($\tau>0$) and $l=0$ if they are not inertial ($\tau=0$). Thus, voters
with opinion $+1$ that changed their opinion in the last update step
would contribute to the quantity $p_0^+(t)$, without an opinion change
they would contribute to $p_1^+(t)$.  The global density of an opinion
$\sigma$ at time $t$ is given by
\begin{equation}
  \label{frequ}
  P_\sigma(t)= p_0^\sigma(t) + p_1^\sigma(t) .
\end{equation}
Fig.~\ref{pst} illustrates the possible transitions of voters from one fraction
to another.
\begin{center}
\vspace{-0.5cm}
 \begin{figure}[ht!]
   \centering
   \psset{unit=1.3cm}
   \begin{pspicture}(-.2,-.5)(3,3)
     \rput(.5,2){\circlenode{a}{$p_0^+$}}
     \rput(2.5,2){\circlenode{b}{$p_1^+$}}
     \rput(.5,.5){\circlenode{c}{$p_0^-$}}
     \rput(2.5,.5){\circlenode{d}{$p_1^-$}}
     \ncarc{->}{a}{c} \ncarc{->}{c}{a}
     \ncarc{->}{a}{b} \ncarc{->}{c}{d}
     \ncarc{->}{d}{a} \ncarc{->}{b}{c}
     \nccircle{->}{b}{.3} \nccircle{->}{d}{.3}
   \end{pspicture}
   \vspace{-0.5cm}
     \caption{Illustration of the four fractions $p_l^\sigma$ and the
     possible transitions of a voter.
}
\label{pst}
\end{figure}
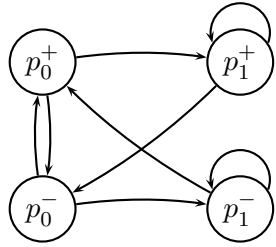
\end{center} 

For the mean field limit, the evolution equations have the form
\begin{eqnarray}
  p_0^\sigma(t+1)-p_0^\sigma(t)&=&W(\sigma|-\sigma,0)p_0^{-\sigma}+W(\sigma|-\sigma,\nu)p_1^{-\sigma}\nonumber\\
& & - (W(-\sigma|\sigma,0)+W(\sigma|\sigma,0))p_0^\sigma \\
p_1^\sigma(t+1)-p_1^\sigma(t)&=&W(\sigma|\sigma,0)p_0^{\sigma}-W(-\sigma|\sigma
,\nu)p_1^\sigma.
\end{eqnarray}
In these equations, the global changing and sticking probabilities are
easily found by using Eqs.~(\ref{lvm}) and (\ref{ivm}),
\begin{eqnarray*} 
W(-\sigma|\sigma,0) &=& W^{V}(-\sigma) =  P_{-\sigma}(t),\\
W(\sigma|\sigma,0) &=& W^{V}(\sigma) = P_{\sigma}(t),\\
W(-\sigma|\sigma,\nu) &=& (1-\nu)\,W^{V}(-\sigma) = (1-\nu)\,P_{-\sigma}(t),\\
W(\sigma|\sigma,\nu) &=& 1 - (1-\nu)\,W^{V}(\sigma) = P_{-\sigma}(t) + \nu
P_{\sigma}(t) .
\end{eqnarray*}
After some steps of straightforward algebra, the former
expressions can
be written in the example of the +1 opinion as
\begin{eqnarray}
  \label{fracs}
  p_0^+(t+1)-p_0^+(t) & = & P_+(t)\big[p_0^-(t)+(1-\nu) p_1^-(t)\big]-p_0^+(t),\\
  \label{fracs1}
  p_1^+(t+1)-p_1^+(t) & = & P_+(t)p_0^+(t)+P_-(t)\, p_1^+(t)(\nu - 1 ),
\end{eqnarray}
and the equivalent terms are found for opinion $-1$.

The global density of the +1 opinion evolves as the sum of
Eqs.~(\ref{fracs}) and~(\ref{fracs1}) 
which yields, after some more straightforward algebra, the change in the
global density
\begin{equation}
  \label{glo+}
   P_+(t+1) - P_+(t) = \nu \left[p_0^-(t)\,p_1^+(t)-p_0^+(t)p_1^-(t)\right].
\end{equation}
For $\nu=0$, i.e. the standard VM, we obtain the general conservation of
magnetization that we already have seen in Eq.~(\ref{vm_mf}). For $\nu>0$
everything depends on the quantities $p_l^\sigma(t)$. If there is no
heterogeneity of social inertia in the system, i.e.  if at some time
either $p_0^+(t)+p_0^-(t)=1$ or $p_1^+(t)+p_1^-(t)=1$, then there also is
no dynamics in the magnetization.  The same holds if both products in the
squared brackets of Eq.~(\ref{glo+}) are equally high.  This is true if
$P_+=P_-$ and the ratio of inertial voters is the same within the two
global densities, i.e. if $p_0^+(t)=p_0^-(t)$.

In the remaining configurations of these four quantities, there is a
dynamics in the magnetization of the system. This implies that even if
the global densities of the opinions are the same ($P_+=P_-=0.5$), we can
find a evolution towards full consensus at one of the opinions.
Interestingly, the opinion whose density is increasing can be the {\em
  minority} opinion in the system. In general, at every timestep an
opinion $\sigma$ has an increasing share of voters in the system whenever
its internal ratio of inertial voters reaches the inequality
\begin{equation}
  \label{inequ}
  \frac{p_1^\sigma}{p_0^\sigma}>\frac{p_1^{-\sigma}}{p_0^{-\sigma}}.
\end{equation}
However, the complete process is nonlinear and, therefore, it is not
possible to derive the final outcome of the dynamics from Eq.
\ref{inequ}.

Note that condition~(\ref{inequ}) is evidence of the important role of
the heterogeneity of voters on the dynamics in the system. More
precisely, the main driving force of the observed ``slower-is-faster''
effect is the voters' heterogeneity with respect to their inertia.

In order to have an analytical estimation of the effect of social inertia
on the times to consensus, we initialize the system in a situation just
after the symmetry is broken. In particular, we artificially set the
initial densities to differ slightly, i.e. we set $p_0^+(0)=1/2+N^{-1}$
and, hence, $p_0^-(0)=1/2-N^{-1}$.\footnote{We also calculated the
  theoretical predictions for breaking the symmetry in the other way,
  namely by setting $p_0^+(0)=1/2-N^{-1}$ and $p_1^+(0)=N^{-1}$. Here,
  again opinion +1 is favored, but this time just by a higher fraction of
  inertial voters. The initial densities of opinions are equal.
  Qualitatively, this procedure leads to the same theoretical
  predictions.} Then we iterate according to Eqs.~(\ref{fracs})
and~(\ref{fracs1}). Furthermore, we assume that the consensus is reached
whenever for one opinion $p_0^\sigma(t) + p_1^\sigma(t) \leq N^{-1}$
holds.\footnote{With the described initial condition, +1 can be the only
  consensus opinion.}  This is due to the fact that for a system of size
$N$, if the frequency of the minority state falls below $N^{-1}$, the
absorbing state is reached.  As we are interested in the effect of
different inertia-levels, we again use $\nu$ as control parameter and
compare the results with computer simulations of the identical setup of
our inertial VM.  In Fig.~\ref{num}, the lines correspond to this
theoretical analysis, where a qualitative agreement can be seen with the
simulation results.

\section{Conclusions\label{conclusion}}

The time for reaching a fully ordered state in a two-state system such as
the voter model is a problem that attracted attention from different
fields in the last years.  In this paper, we studied the effect of social
inertia in the VM based on the assumption that social inertia grows with
the time the voter has been keeping its current opinion. We focus our
study on how the times to consensus vary depending on the level of
inertia in the population ($\nu$).

Counter-intuitively to the expectation that increasing inertia may lead
to increasing times to reach consensus, we find that, for intermediate
values of $\nu$, this inertia mechanism causes the system to reach
consensus \emph{faster} than in the standard VM. We show that this
phenomenon is robust against the exact topology of the neighborhood
network as we find it in regular lattices and small-world networks. In
the former holds that the larger the dimension, the more noticeable the
effect. Furthermore, we found that the phenomenon also appears in random
and fully-connected networks.

In simple words, this intriguing effect can be understood as follows: Due
to fluctuations, one of the opinions is able to acquire a slight majority
of voters. Therefore, voters of this opinion change less likely and,
hence, the average inertia of this opinion will be higher than the
other. Since inertia reduces emigration, but not immigration, the
majority will become even larger. This development is enforced by higher
values of $\nu$ and constitutes a clear direction of the ordering
dynamics which intuitively can lead to a faster reaching of
consensus. However, for high values of $\nu$, this development is
outperformed by the high level of average inertia in the complete system,
i.e. also within the minority population of voters, which slows down the
overall time scale of the ordering dynamics (compare Fig.~\ref{fixed}).

Interestingly, this phenomenon implies that individuals reluctant to
change their opinion can have a counter-intuitive effect on the consensus
process, which was studied for some particular cases
before~\cite{galam2005}.  Furthermore, an inertial minority can overcome
a less inertial majority in a similar fashion as previously discussed
in~\cite{galam, sanmiguel}.

Whereas, in a recent paper, we derived the complete macroscopic dynamics
of a system with slowly increasing inertia~\cite{PRL}, here we discuss a
reduced model based on only two levels of inertia. Albeit simple, this
model can still give rise to the observed ``slower-is-faster''
phenomenon. It also allows for a theoretical approach to unveil its
origin, namely the described ageing-mechanism that breaks the
magnetization conservation. This is different from the standard VM where
magnetization is always conserved. We showed that the break of
magnetization conservation only holds when the voters build up a {\em
  heterogeneity} with respect to their inertia to change opinion.
Therefore, once the symmetry between (a) the global densities of the two
possible states and/or (b) the proportions of inertial voters is broken,
the favored state (opinion) achieves both (i) a reinforcement of its
average inertia and (ii) a fast recruitment of the less inertial state.
Both effects contribute to a faster deviation from the symmetric state.
For some parameter ranges, these mechanisms outweigh the increasement in
the time to reach consensus generated by the high inertia of the state
that disappeared in equilibrium.

\section*{Acknowledgements}
HUS wants to thank Stefano Battiston, Matthias Feiler, and Patrick
Gr\"ober for many fruitful discussions and useful hints, and Frank E.
Walter for technical support. CJT acknowledges financial support of SBF
(Switzerland) through research project C05.0148 (Physics of Risk).


\begin{thebibliography}{10}

\bibitem{abelson64mmd}
R.P. Abelson.
\newblock {Mathematical models of the distribution of attitudes under
  controversy}.
\newblock {\em Contributions to Mathematical Psychology}, 14:1--160, 1964.

\bibitem{castellano2003}
C.~Castellano, D.~Vilone, and A.~Vespignani.
\newblock Incomplete ordering of the voter model on small-world networks.
\newblock {\em Europhysics Letters}, 63:153--158, 2003.

\bibitem{xavi}
Xavier Castell\`o, Victor~M. Egu\'{\i}luz, and Maxi San~Miguel.
\newblock Ordering dynamics with two non-excluding options: bilingualism in
  language competition.
\newblock {\em New Journal of Physics}, 8:308, 2006.

\bibitem{chave2001}
J.~Chave.
\newblock Spatial patterns and persistence of woody plant species in ecological
  communities.
\newblock {\em The American Naturalist}, 157:51--65, 2001.

\bibitem{deffuant00mba}
G.~Deffuant, D.~Neau, F.~Amblard, and G.~Weisbuch.
\newblock {Mixing beliefs among interacting agents}.
\newblock {\em Advances in Complex Systems}, 3(1-4):87--98, 2000.

\bibitem{dornic2001}
I.~Dornic, H.~Chat\'e, J.~Chave, and H.~Hinrichsen.
\newblock Critical coarsening without surface tension: The universality class
  of the voter model.
\newblock {\em Physical Review Letters}, 87(4):045701, Jul 2001.

\bibitem{dugatkin01}
Lee~Alan Dugatkin.
\newblock {\em The imitation factor: evolution beyond the gene}.
\newblock Free Press, New York, 2000.

\bibitem{fontes2001}
L.~R. Fontes, M.~Isopi, C.~M. Newman, and D.~L. Stein.
\newblock Aging in 1d discrete spin models and equivalent systems.
\newblock {\em Physical Review Letters}, 87(11):110201, Aug 2001.

\bibitem{frachenbourg1996}
L.~Frachebourg and P.~L. Krapivsky.
\newblock Exact results for kinetics of catalytic reactions.
\newblock {\em Physical Review E}, 53(4):R3009--R3012, Apr 1996.

\bibitem{french56fts}
J.R.P. French.
\newblock {Formal Theory of Social Power}.
\newblock {\em Psychological Review}, 63:181--194, 1956.

\bibitem{galam2005}
S.~Galam.
\newblock Local dynamics vs. social mechanisms: A unifying frame.
\newblock {\em Europhysics Letters}, 20:705, 2005.

\bibitem{galam}
Serge Galam.
\newblock Minority opinion spreading in random geometry.
\newblock {\em European Physical Journal B}, 25:403--406, 2002.

\bibitem{gunton1983}
J.~D. Gunton, M.~San~Miguel, and P.~S. Sahni.
\newblock {\em Phase Transitions and Critical Phenomena}.
\newblock Academic Press, London, 1983.
\newblock C. Domb and J. Lebowitz, Eds.

\bibitem{hegselmann02oda}
R.~Hegselmann and U.~Krause.
\newblock {Opinion Dynamics and Bounded Confidence Models, Analysis, and
  Simulation}.
\newblock {\em Journal of Artifical Societies and Social Simulation (JASSS)},
  5(3), 2002.

\bibitem{helbing}
D.~Helbing, I.~Farkas, and T.~Vicsek.
\newblock Simulating dynamical features of escape panic.
\newblock {\em Nature}, 407:487--490, 2000.

\bibitem{holley75}
Richard~A. Holley and Thomas~M. Liggett.
\newblock Ergodic theorems for weakly interacting infinite systems and the
  voter model.
\newblock {\em The Annals of Probability}, 3(4):643--663, 1975.

\bibitem{holyst2001}
J.A. Ho{\l}yst, K.~Kacperski, and F.~Schweitzer.
\newblock {Social impact models of opinion dynamics}.
\newblock {\em Annual Reviews of Computational Physics}, 9:253--273, 2001.

\bibitem{ligget1995}
T.~M. Liggett.
\newblock {\em Interacting Particle Systems}.
\newblock Springer, New York, 1995.

\bibitem{lorenz07cod}
J.~Lorenz.
\newblock {Continuous Opinion Dynamics under Bounded Confidence: A Survey}.
\newblock {\em International Journal of Modern Physics C}, 18(12):1--20, 2007.

\bibitem{maslov2003}
S.~Maslov, K.~Sneppen, and U.~Alon.
\newblock Correlation profiles and motifs in complex networks.
\newblock In S.~Bornholdt and H.G. Schuster, editors, {\em Handbook of graphs
  and networks. {F}rom the genoma to the internet}, pages 169--198. Wiley VCH
  and Co., 2003.

\bibitem{ravasz04}
Maria Ravasz, Gy\"orgy Szabo, and Attila Szolnoki.
\newblock Spreading of families in cyclic predator-prey models.
\newblock {\em Physical Review E (Statistical, Nonlinear, and Soft Matter
  Physics)}, 70(1):012901, 2004.

\bibitem{redner2001}
S.~Redner.
\newblock {\em A guide to first-passage processes}.
\newblock Cambridge University Press, Cambridge, 2001.

\bibitem{fs-wcss-06}
Frank Schweitzer.
\newblock Collective decisions in multi-agent systems.
\newblock In S.~Takahashi, D.~Sallach, and J.~Rouchier, editors, {\em Advancing
  Social Simulation. The First World Congress}, pages 7--12. Springer, Tokyo,
  2007.

\bibitem{slanina2003}
F.~Slanina and H.~Lavicka.
\newblock {Analytical results for the Sznajd model of opinion formation}.
\newblock {\em European Physical Journal B}, 35(2):279--288, 2003.

\bibitem{sood2005}
V.~Sood and S.~Redner.
\newblock Voter model on heterogeneous graphs.
\newblock {\em Physical Review Letters}, 94(17):178701, 2005.

\bibitem{PRL}
H.-U. Stark, C.~J. Tessone, and F.~Schweitzer.
\newblock Decelerating microdynamics can accelerate macrodynamics in the voter
  model.
\newblock {\em Physical Review Letters}, 101(1):018701(4), 2008.

\bibitem{suchecki2005}
K.~Suchecki, V.~M. Egu\'{\i}luz, and M.~San~Miguel.
\newblock Conservation laws for the voter model in complex networks.
\newblock {\em Europhysics Letters}, 69:228--234, 2005.

\bibitem{maxi2}
K.~Suchecki, V.~M. Egu\'{\i}luz, and M.~San~Miguel.
\newblock Voter model dynamics in complex networks: Role of dimensionality,
  disorder, and degree distribution.
\newblock {\em Physical Review E}, 72:036132, 2005.

\bibitem{sanmiguel}
C.~J. Tessone, R.~Toral, P.~Amengual, H.~S. Wio, and M.~San~Miguel.
\newblock Neighborhood models of minority opinion spreading.
\newblock {\em European Physical Journal B}, 39:535--544, 2004.

\bibitem{watts1998}
D.J. Watts and S.H. Strogatz.
\newblock Collective dynamics of small-world networks.
\newblock {\em Nature}, 393:440, 1998.

\end{thebibliography}
\end{document}